\documentclass[sigconf]{acmart}

\setcopyright{acmlicensed}
\copyrightyear{2025}
\acmYear{2025}
\acmDOI{XXXXXXX.XXXXXXX}
\acmConference[EASE 2025]{The 29th International Conference on Evaluation and Assessment in Software Engineering}{17–20 June, 2025}{Istanbul, Türkiye}

\acmISBN{978-1-4503-XXXX-X/18/06}

\acmSubmissionID{126}

\usepackage{colortbl}
\usepackage{cleveref}
\usepackage[group-separator={,}]{siunitx}
\usepackage{enumitem}
\newcommand{\etal}{\textit{et al}.}
\newcommand{\ie}{\textit{i}.\textit{e}.}

\definecolor{lblue}{HTML}{A6CEE3}
\begin{document}

\title{On the Prevalence and Usage of Commit Signing on GitHub}
\subtitle{A Longitudinal and Cross-Domain Study}

\author{Anupam Sharma}
\email{sharmaanupam@iitgn.ac.in}
\orcid{0000-0002-3443-4646}
\affiliation{%
  \institution{Indian Institute of Technology Gandhinagar}
  \city{Gandhinagar}
  \state{Gujarat}
  \country{India}
}
\author{Sreyashi Karmakar}
\email{24310059@iitgn.ac.in}
\orcid{0000-0002-5578-0396}
\affiliation{%
  \institution{Indian Institute of Technology Gandhinagar}
  \city{Gandhinagar}
  \state{Gujarat}
  \country{India}
}
\author{Gayatri Priyadarsini Kancherla}
\email{gayatripriyadarsini@iitgn.ac.in}
\orcid{0000-0002-1842-9353}
\affiliation{%
  \institution{Indian Institute of Technology Gandhinagar}
  \city{Gandhinagar}
  \state{Gujarat}
  \country{India}
}
\author{Abhishek Bichhawat}
\email{abhishek.b@iitgn.ac.in}
\orcid{0000-0002-3075-2743}
\affiliation{%
  \institution{Indian Institute of Technology Gandhinagar}
  \city{Gandhinagar}
  \state{Gujarat}
  \country{India}
}




\begin{abstract}
GitHub is one of the most widely used public code development platform. However, the code hosted publicly on the platform is vulnerable to \emph{commit spoofing} that allows an adversary to introduce malicious code or commits into the repository by spoofing the commit metadata to indicate that the code was added by a legitimate user. The only defense that GitHub employs is the process of commit signing, which indicates whether a commit is from a valid source or not based on the keys registered by the users.

In this work, we perform an empirical analysis of how prevalent is the use of commit signing in commonly used GitHub repositories. To this end, we build a framework that allows us to extract the metadata of all prior commits of a GitHub repository, and identify what commits in the repository are verified. 
We analyzed 60 open-source repositories belonging to four different domains — web development, databases, machine learning packages and security — using our framework and study the presence of verified commits in each of these repositories over five years. 
Our analysis shows that only $\sim$10\% of all the commits in these 60 repositories are verified. 
Developers committing code to security-related repositories are much more vigilant when it comes to signing commits by users. 

We also analyzed different Git clients for the ease of commit signing through their interfaces, and found that GitKraken provides the most convenient way of commit signing whereas GitHub Web provides the most accessible way for verifying commits. During our analysis, we also identified an unexpected behavior in how GitHub handles unverified emails in user accounts preventing legitimate owner to use the email address. 
We believe that the low number of verified commits may be due to lack of awareness, complicated steps for setup, and difficulty in managing multiple keys across systems. Finally, we propose ways to identify commit ownership based on GitHub's Events API addressing the issue of commit spoofing.
\end{abstract}



\maketitle

\section{Introduction}
Version control systems have become a standard tool for software development where developers worldwide can collaborate on a project. Git~\cite{git} is one such widely used version control system, and GitHub~\cite{github} is a commonly used repository-hosting and collaborative software development platform based on Git. Multiple Git clients like Git-CLI~\cite{git}, GitHub Desktop~\cite{github-desktop} and GitKraken~\cite{git-kraken} allow users to work on projects or repositories being developed on GitHub. Public repositories or projects on GitHub can be downloaded or cloned by all users while changes to the project are either performed through pull requests or by directly committing to the repository. Although owners or managers of a repository review the pull requests, it is possible that a malicious user can inject unintended code into the repository without getting flagged or noticed by them. One such method of injecting malicious code is via \emph{commit spoofing} where an attacker alters the metadata of commit to impersonate another user and introduce malicious commits. For instance, in 2021, malicious commits were pushed to the \texttt{php-src} repository using the names of legitimate users~\cite{git-php-issue}. Attackers can use commit spoofing to perform typo-squatting, where attackers create clones of legitimate repositories with similar names and introduce malicious commits~\cite{whatthefork}. Users can get tricked due to similar names, and may download the malicious repository.


The origin of commits is identified using the username and email associated with the commit. However, GitHub associates the commit with the user account in GitHub without any verification. A mere presence of email address in the commit data is enough to associate the commit with an account having the same email address. As Git allows updating Git configuration with any email address, it is possible for malicious actors to push code on Git and make it appear as being committed by some other user, \ie, a malicious user can set their Git configuration to mimic the username and email of a legitimate or known user, and may be able to commit to repositories without being identified. This attack was successfully performed, recently, by unknown users on open-source libraries~\cite{github-attack,git-php-issue}. It is, in fact, quite easy to exploit this feature and perform commit spoofing in repositories~\cite{commit-attack}. 

GitHub introduced verification badges and vigilant mode to prevent spoofing and provide guarantees that the commits originated from a trusted source~\cite{vigilantMode}. Commit signing uses digital signatures to ensure that every commit on GitHub is signed using the key registered by the user in their GitHub account. Commits displayed on the GitHub page, with vigilant mode turned on, are then tagged with labels or \emph{badges} depending on the verification of the user commits. 
While this is a useful feature to warn users about potentially malicious commits, it is unclear if the developers use this feature to sign their commits on GitHub repositories, and if so how prevalent it is. The feature itself is available only on the \texttt{commits} page of repositories in GitHub, which is rarely accessed by the users. Moreover, it is also not known if popular Git clients assist developers by providing user-friendly features to perform commit signing and verification. We try to address both these issues in the current work. More specifically, we address the following research questions:
\begin{itemize}
    \item[\textbf{RQ1:}] \textit{How prevalent is the usage of commit signing among the developers of different open-source repositories on GitHub?}
    \begin{enumerate}
        \item How often are the features against commit spoofing, provided by GitHub, used in practice in popular open-source repositories?
        \item What are the usage trends of commit signing before and after the release of features like verification badges and vigilant mode by GitHub?
        \item Does the background or domain of a user impact their commit signing practices? More concretely, do a certain category of developers, say web-developers, use commit signing more often than other category of developers, say, security developers?
    \end{enumerate}
    \item[\textbf{RQ2:}] \textit{Do popular Git clients like Git CLI, GitHub Desktop, and GitKraken, in comparison to GitHub web, implement convenient features to perform commit signing or have any provisions against commit spoofing, and if the existing states of these software can be used against commit spoofing?} 
\end{itemize}

To this end, we developed a framework\footnote[1]{The code and the dataset is at \url{https://github.com/anp-scp/commit_crawler}} to analyze the commits of a given repository. Given a GitHub repository, the framework fetches commit metadata based on the date range provided and provides distribution of commits as verified and unverified commits based on the signature attached with the commits. We performed a longitudinal study analyzing the commits on 60 top repositories (based on the Gitstar Ranking~\cite{gitStarRank}) from four domains (15 repositories from each domain) --- machine learning, security, database and web development --- over five years, \ie, one year before and four years after the vigilant mode and verification badges were introduced by GitHub. We observed that only $\sim10\%$ of the total commits were signed and the usage of signed commits increased in the first year after the introduction of vigilant mode by GitHub, but later decreased constantly. Within each of the four domains, 12.77\% of commits in web development repositories, 2.55\% of commits in ML-related repositories, 5.95\% of commits in database-related repositories and 28.35\% of commits in security-related repositories were verified --- of the four domains, security developers tend to use commit signing the most. We also found that, of the three popular Git clients --- Git CLI, GitHub Desktop and GitKraken, GitKraken provides the most convenient way of commit signing whereas GitHub web provides the most accessible way for verifying commits. 
We discuss these results in detail in \cref{sec:results1,sec:discussion}. 

During the analysis, we also discovered an unexpected behavior in how GitHub handles "unverified" emails in commit metadata. This behavior allows an attacker to add any email address, which is not already in use on GitHub, to their account and push commits using that email address.
GitHub, then, associates the commit with the attacker's account even if the email address is unverified. The email address remains linked to the attacker, preventing the rightful owner from using it with GitHub in the future. Moreover, the verification mail sent to the actual owner provides no means of reporting the incident and instructs the owner to ignore the mail if the owner did not initiate the action, failing to raise any alarm. While the owner can claim the email address by contacting GitHub support, this is possible only when the owner becomes aware of the misuse. We discuss this in detail in \Cref{bug,disc:bug}. We disclosed this behavior to GitHub; their response is available in our code repository\footnotemark[1].

In summary, our key contributions are as follows:
\begin{itemize}
    \item We performed a longitudinal and cross-domain analysis of the prevalence of commit signing. Unlike existing studies, we studied the commits of actual users by excluding the commits made via bots and GitHub web. 
    \item We assessed the usability of GitHub's web interface, Git CLI, GitHub Desktop and GitKraken from the perspective of commit signing and verification. 
    \item We also discovered an unexpected behavior in how GitHub handles unverified emails in user accounts and GitHub, preventing legitimate owner to use the email address. 
    \item Lastly, we provide recommendations, in \cref{disc:alternate}, on alternate ways of associating commits with user accounts in GitHub that could help defend against commit spoofing. 
\end{itemize}
\section{Background}

\subsection{Commit Spoofing}

\begin{figure*}[h]
  \centering
  \includegraphics[width=0.88\linewidth]{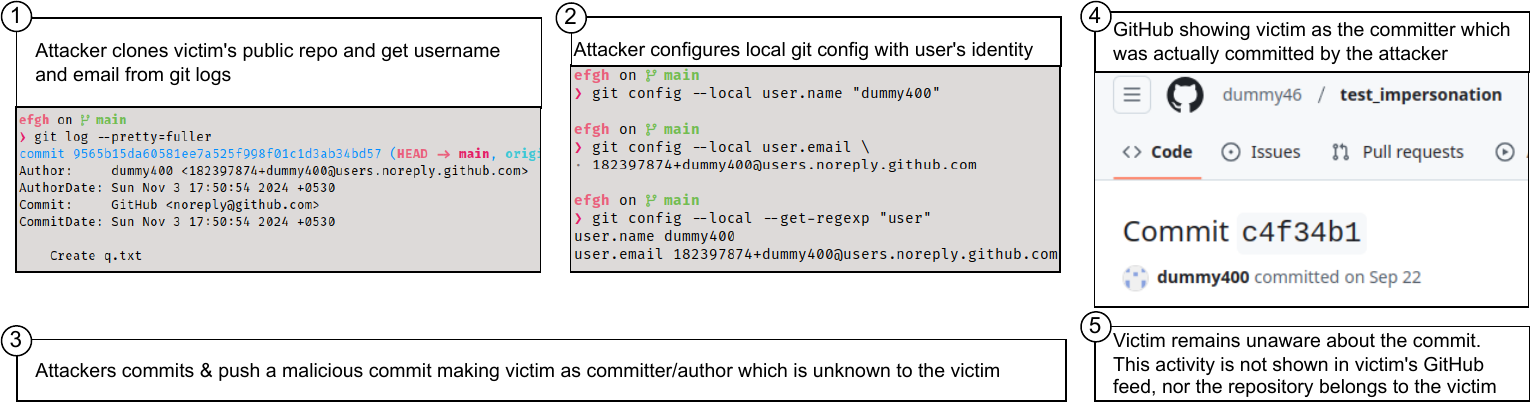}
  \caption{Commit spoofing. The attacker (\texttt{dummy46}) gets the identity information from the victims's (\texttt{dummy400}) public repository logs and uses it in the Git config to push malicious commits on victim's behalf. GitHub associates the commit to the victim.}
  \Description{The impersonator (dummy46) gets the identity information from the user's public repository (dummy400/efgh) logs and uses it in the Git configuration to push malicious commits in the user's name in the repository named ``dummy46/test\_impersonation''. GitHub associates the commit to the user even though the impersonator committed it.}
  \label{fig:impersonationFlow}
\end{figure*}

\begin{figure}[!tbp]
  \centering
  \fbox{\includegraphics[width=\linewidth]{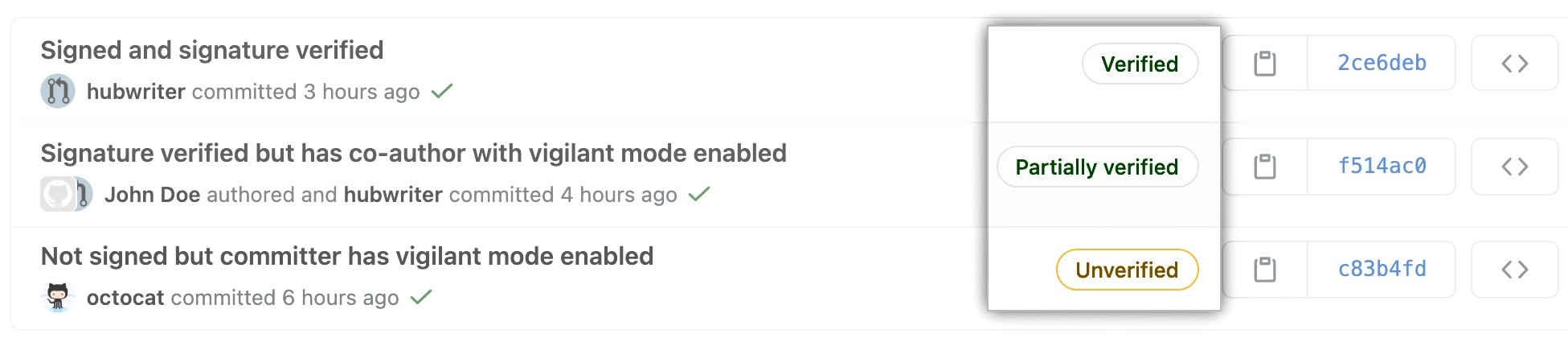}}  
  \caption{The different verification badges assigned by GitHub based on commit signatures~\cite{vigilantMode}.}
  \label{fig:Badges}
\end{figure}

Commit spoofing is a form of identity theft where a developer can alter the metadata of a Git commit to record that the commit originated from another user.
This spoofing takes advantage of two facts: (i) Git uses user-provided metadata to track the author of the commit, such as the author’s name and email address, without any identity verification, and (ii) GitHub uses the provided email address to associate the commit with a GitHub user account. An attacker can easily get the email address from someone's public repository either via \texttt{git log} or via the \texttt{patch} format from GitHub web, configure a repository with the obtained identity information, and commit code to GitHub. This creates a false impression that the victim has authored the code, which can harm the reputation of the victim, resulting in defamation for the victim, or inject vulnerabilities in a repository via a pull request, tricking maintainers into thinking that it came from a legitimate user.

\Cref{fig:impersonationFlow} shows the steps involved in performing commit spoofing. The GitHub account of the attacker is \texttt{dummy46} and the victim's account is \texttt{dummy400}. The attacker (\texttt{dummy46}) first gets the identity information from the Git log of a public repository (\textcircled{1}), configures the local Git config with the victim's identity (\textcircled{2}), and uses it to commit and push code to a repository unknown to the user (\textcircled{3}). Since GitHub uses email addresses to associate commits with user accounts, GitHub shows the profile of the victim (\texttt{dummy400}) as the author of the commit (\textcircled{4}). The victim remains completely unaware of the commit (\textcircled{5}). The commit is neither visible in the attacker's activity overview nor in the victim's activity overview in their respective GitHub profiles, preventing the verification of the authorship of the commit.

\subsection{Commit Signing, Badges and Vigilant Mode}
\label{background: badges}
Commit signing is a technique to defend against commit spoofing; using the \texttt{git commit –S} command, commits can be signed using GPG, SSH, or S/MIME keys. In addition to this feature, GitHub released the vigilant mode and verification badges on April 28, 2021 \cite{vigilantMode}. \Cref{fig:Badges} shows the different kinds of verification badges visible on the GitHub web interface, which helps in differentiating commits. All correctly signed commits are labeled with the ``Verified'' badge. The incorrectly signed or unsigned commits are labeled with the ``Unverified'' badge since they may contain illegitimate commits.

In addition to verification badges, GitHub has a vigilant mode feature for users. If a user enables vigilant mode, any unsigned commits associated with the user are marked as ``Unverified''. This prevents commit spoofing because if an attacker tries to spoof the commit metadata with the details of another legitimate user, and if that user's vigilant mode is enabled, then the commit made by the attacker will be automatically marked as ``Unverified''. However, if the vigilant mode is disabled, all unsigned commits are simply unlabeled, \ie, not given any badge. But, the disadvantage of this feature is all the legitimate commits that remained unsigned due to lack of awareness will also be marked as ``Unverifed".

Additionally, GitHub displays a ``Partially Verified'' badge, where a commit has multiple authors, as permitted by Git. For instance, a committer can set another user as the co-author using \texttt{git commit --author} command, and one can perform spoofing by taking advantage of this feature. If there exists such a commit in GitHub, and the user marked as the author has enabled the vigilant mode, then the commit is marked as ``Partially Verified''; otherwise, it is simply marked as ``Verified''. Hence, vigilant mode provides an additional layer of security to maintain the integrity of commits as, if someone adds other authors in a commit and the commit is signed, GitHub will mark the commit as ``Partially Verified'' alerting the users that not all authors have signed the commit.

\section{Related Works}

Git is the most widely used modern version control system, with GitHub alone hosting over 100 million developers and more than 420 million repositories \cite{githubBuildSoftware}. In this section, we review relevant research in this domain. We begin by discussing various attacks that have been explored, followed by a detailed examination of prior work on commit spoofing. Finally, we conclude with a discussion on quantitative and empirical studies related to commit signing, providing insights into the current landscape.\\

\subsection{Attacks and Vulnerabilities on GitHub}
Attacks targeting GitHub's functionality can be broadly classified into three categories: metadata manipulation attacks, integrity attacks (code manipulation), and commit spoofing.

Torres-Arias \etal~\cite{197135} discuss several attacks targeting Git's metadata, which can lead to inconsistent repository states and unintended developer actions. These attacks exploit vulnerabilities in Git's handling of metadata, such as branch and tag references, rather than the content of the commits themselves. These attacks are subtle and can leave no trace, making them particularly dangerous. They primarily target the metadata that Git uses to organize and reference commits instead of the using the commit objects. The authors propose maintaining a cryptographically-signed log of relevant developer actions. The exploit leveraged a weakness that allowed attackers to subtly manipulate Git repository structures by targeting unprotected metadata. The authors propose using cryptographic mechanisms to safeguard the data. 

Another category of attacks studied by previous works include man-in-the-middle attack, which eventually lead to integrity violations. The work done by Afzali \etal~\cite{Afzali_legitimate} discusses such scenarios~\cite{greatfireChinaGitHub, china_great_cannon} where a legitimate user utilizes the web interface of the code hosting service to commit a code, and the hosting service itself is compromised. When a user commits code from GitHub Web, the commit is actually made by GitHub on behalf of the user. In this case, the commit's author is the user, and GitHub is the committer. GitHub will sign the commit by default and show it as verified. However, if the Git server is compromised, an attacker can tamper with the changes requested by the user, and GitHub will still show it as verified. For instance, a user was able to upload their public key to the Rails project by exploiting a vulnerability in public key update form in GitHub~\cite{githubPublicSecurity, githubComeCommit}. The solution is to sign the commit, which makes any modification unverified, assuming the attacker does not have the private key. The authors proposed a browser extension to perform commit signing while performing web-based commit. Such attacks apply to all Git and web-based code hosting software, specifically in an on-premise setup. It is important to note that commit signing prevents such impersonation unless the GPG web-of-trust is not breached. If web-of-trust is breached, the attacker can still perform such impersonation. Hence, it is essential to study the proper setup of web-of-trust in the Git ecosystem.

\subsection{Commit Spoofing}
Git allows viewing the authors' and committers' email via the \texttt{git log} command. Similarly, GitHub allows the viewing of the code author's email from the patch view of a commit. One can access the patch view by adding \texttt{.patch} at the end of a commit URL of a public GitHub repository (to get the commit URL, one can open the commit history of a GitHub repository and click on the hash of any commit). An attacker can use these details in the Git configuration file to impersonate someone else by committing and pushing code to a GitHub repository. Since GitHub uses email to associate users in web interface, the victim's profile would be shown by GitHub as the author of the code without the victim's knowledge. Such attacks can have highly defamatory implications for the victim; hence, studying the defenses against such impersonation is important.

Another consequence of such impersonation is malicious clones of reputed repositories with slight organizational and repository name changes (typo-squatting). Another level of trust can be built by adding stars from fake accounts and committing code using the email of the maintainer of the legitimate repository. These steps can easily trick users into thinking it is a legitimate repository. Cao \etal~\cite{whatthefork} devised a system, Fork Sentry, to detect malicious forks. The authors identified 26 malicious forks used for illicit mining out of 35 cryptocurrency repositories. 
The identification could have also been performed by checking for signed commits and verification badges as the attacker will never be able to get a verified badge on the impersonated commit, if the victim has enabled the vigilant mode. In a similar work by Gonzalez \etal~\cite{anomalicious}, the authors curated a dataset of 15 malware-infected repositories in which their rule-based decision model was able to flag 53.33\% of malicious commits. The authors considered factors like lines of code added, deleted, and modified, number of files added, deleted, or modified, sensitive files edited, file history, and contributor's trust. 

Bottolfsen \etal~\cite{bottolfsen2024investigation} discuss the importance of improving security practices in software development to mitigate the risks associated with commit spoofing. This work discusses the implication of commit spoofing and the corresponding mitigation strategies. They, additionally, propose a tool which identifies a commit spoofing in repositories, when a code is pushed or merged, via a GitHub action. However, it cannot flag any of the commits that were made before the tool is installed and doesn't handles the scenario where a repository is cloned, updated and then pushed. We discuss our approach considering all such scenarios in detail in \cref{sec:discussion}.     


\subsection{Measurement and Usability Studies}
Research on GitHub security often targets the accidental exposure of sensitive information within repositories. This involved secret strings extraction from repositories through various measurement studies~\cite{10.1145/3510003.3510150,inproceedings_ndss} where the strings identified to be secrets include API keys, cryptographic keys, and other authentication credentials, which are critical for security. 
Prior studies~\cite{10.1145/2597073.2597117,10.1145/3350546.3352519} identified vulnerabilities in open-source GitHub repositories whose issues are mined from discussions around commits and pull requests. Other sources include social media platforms like X (formerly, Twitter), Reddit, and online blogs and articles.

Recently, a quantitative study by Collier~\cite{collier2024quantitative} analyzed GitHub users' profile to determine the prevalence of signed commits, concluding that commit signing remains an uncommon practice. While our study shares a similar objective, we focus on repositories rather than individual users. This allows us to better capture the variability of commit signing practices within a given repository. Additionally, we analyze the longitudinal trend of signed commits, assessing changes before and after the introduction of GitHub's vigilant mode to evaluate its impact on adoption. Similarly, Bottolfsen~\cite{bottolfsen2024investigation} conducted a quantitative analysis examining factors correlated with signed commits in repositories, including programming language, repository size, and number of forks. Our analysis further investigates commit signing across different repository categories to understand which developer communities demonstrate greater awareness of commit signing security practices.
Unlike previous studies, we exclude commits made by automated bots, which could otherwise skew the data about human developers' commit signing practices. 

Several studies~\cite{8886906,4140996,article,inproceedings2} have highlighted key mismanagement and the challenges of digital signature usability as critical failure points. Research on SSH key systems~\cite{10.1007/978-3-540-25980-0_4,article2} has examined methods to enhance security layers, improve key management, and optimize user interactions. While prior studies have investigated user experiences with secure emails and SSH key setup in general practice, a comprehensive usability study of commit signing on GitHub remains unexplored. In this study, we analyze different Git client interfaces, assess the usability of vigilant mode, and examine the various badges and tags displayed in different scenarios. 





\section{Study Methodology and Setup}

\begin{table*}[]
\caption{Top 15 repositories related to tools or libraries having at least 1K commits were considered for the study based on their popularity in Gitstar Ranking \cite{gitStarRank} for each of the 4 categories. This list excluded repositories having less than 20\% of total commits during our extraction interval (28 April 2020 to 31 December 2024).}
\resizebox{0.84\textwidth}{!}{
\rowcolors{1}{white}{gray!25}
\footnotesize
\begin{tabular}{@{}lllll@{}}
\toprule
Sl. No. & Machine Learning                 & Database              & Security                    & Web dev               \\ \midrule
1       & tensorflow/tensorflow            & redis/redis           & fatedier/frp                & facebook/react        \\
2       & AUTOMATIC1111/stable-diffusion-webui & ClickHouse/ClickHouse & NationalSecurityAgency/ghidra & vercel/next.js \\
3       & huggingface/transformers         & pingcap/tidb          & gorhill/uBlock              & facebook/react-native \\
4       & pytorch/pytorch                  & cockroachdb/cockroach & mitmproxy/mitmproxy         & nodejs/node           \\
5       & tensorflow/models                & influxdata/influxdb   & rapid7/metasploit-framework & mrdoob/three.js       \\
6       & keras-team/keras                 & facebook/rocksdb      & gchq/CyberChef              & denoland/deno         \\
7       & scikit-learn/scikit-learn        & surrealdb/surrealdb   & schollz/croc                & angular/angular       \\
8       & geekan/MetaGPT                   & mongodb/mongo         & openssl/openssl             & mui/material-ui       \\
9       & hiyouga/LLaMA-Factory            & duckdb/duckdb         & AdguardTeam/AdGuardHome     & ant-design/ant-design \\
10      & microsoft/autogen                & taosdata/TDengine     & brave/brave-browser         & puppeteer/puppeteer   \\
11      & microsoft/DeepSpeed              & pubkey/rxdb           & cure53/DOMPurify            & tauri-apps/tauri      \\
12      & openai/gym                       & timescale/timescaledb & zaproxy/zaproxy             & storybookjs/storybook \\
13      & ray-project/ray                  & rqlite/rqlite         & cryptomator/cryptomator     & sveltejs/svelte       \\
14      & hankcs/HanLP                     & tikv/tikv             & SoftEtherVPN/SoftEtherVPN   & gin-gonic/gin         \\
15      & huggingface/pytorch-image-models & scylladb/scylladb     & OpenVPN/openvpn             & gohugoio/hugo         \\ \bottomrule
\end{tabular}%
}
\label{tab:repoInfo}
\end{table*}

Our study investigates: (i) how are the defenses provided by GitHub against commit spoofing adopted by the open-source community, specifically among GitHub users, (ii) does the area or domain of the developer have any effect on its adoption, and (iii) if popular Git clients support features to mitigate commit spoofing. We describe our data collection and analysis approaches, next.

\subsection{Data Collection}

\begin{figure*}[!tbp]
  \centering
  \includegraphics[width=0.7\linewidth]{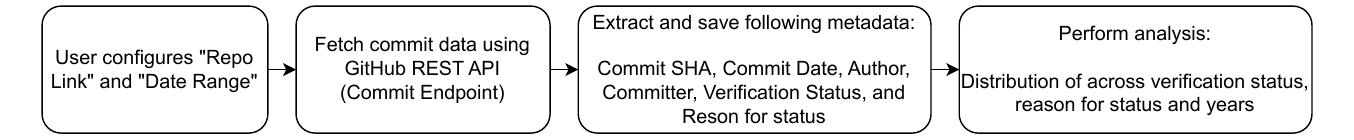}
  \caption{Flow diagram showing the usage of the framework for commit extraction and analysis.}
  \label{fig:framework}
\end{figure*}

\subsubsection{\textbf{Repository Selection}}
We selected repositories from four categories or domains: (i) machine learning, (ii) database, (iii) security, and (iv) web development. We referred to Gitstar Ranking~\cite{gitStarRank} to select the repositories based on their popularity. We iterated through some of the top repositories listed in there and categorized them into the 4 domains based on their description. We excluded repositories that could not be classified under one of these domains, which were not a tool or libraries (for example, tutorials, guides, lists of awesome research papers in some domain, etc) and have less than 1K commits in total at the time of doing the study. Additionally, we excluded repositories that had less than 20\% commits during our extraction interval (the time interval for which we downloaded the commits; we discuss this in the next step) out of the total commits. Finally, we considered 15 top repositories in each of the four domains for our analysis; the list of repositories is shown in \cref{tab:repoInfo}.

\subsubsection{\textbf{Commit Extraction}}

For extracting commit metadata, we utilized GitHub's REST API~\cite{github-rest-api}. However, unlike GitHub's web interface, the REST API marks a commit as either \texttt{verified} or \texttt{unverified} with a reason instead of the different badges.
The REST API marks correctly signed commits as \texttt{verified} and all unsigned and incorrectly signed commits as \texttt{unverified} with possible reasons covering the "no badge" commits, too. The only way to track the badges in the GitHub's web interface is to crawl the commit pages of the repositories. However, we opt for the REST API instead of crawling the web pages as (i) crawling of the commit pages in GitHub is not allowed as per GitHub's \texttt{robots.txt}, (ii) there is a rate limit provided for sending requests to GitHub's REST API server but no such standards are provided for crawling of GitHub pages, and (iii) the case for commits with ``Partially Verified" badge and no badge is dependent on the ``Vigilant Mode" settings of a user in GitHub.

We used the \texttt{commit} endpoint of GitHub's REST API to get the commit-related information of the repositories, which included ``COMMIT SHA'', ``COMMIT\_DATE'', ``AUTHOR'', ``COMMITTER'', ``STATUS'' and ``REASON''. Here, the ``STATUS'' is a boolean that specifies if the commit is verified, and the ``REASON'' is an enum whose value specifies why a commit is assigned the respective status. \Cref{tab:signStatus} describes the possible values for this field. As GitHub released vigilant mode and verification badges on April 28th, 2021~\cite{vigilantMode}, we downloaded the commit data from one year before the release, \ie, April 28th, 2020, until December 31st, 2024, to observe if the commits reflect any change in commit signing practices \emph{before} and \emph{after} the release of vigilant mode and verification badges. We refer to this time period as the ``extraction interval".

We created a framework for the extraction and analysis, which follows the above steps to fetch the commit data given the extraction interval and the repository link as shown in \cref{fig:framework}. We downloaded a total of $869,437$ commits across all 60 repositories in the last five years.

\begin{table}[ht]
\caption{Description of different states of commit based on commit signature as per GitHub REST API \cite{githubCommitRest}}
\resizebox{0.9\columnwidth}{!}{%
\rowcolors{1}{white}{gray!25}
\begin{tabular}{@{}ll@{}}
\toprule
Status & Description \\ \midrule
valid & \begin{tabular}[c]{@{}l@{}}The signature is regarded as valid as none of the below\\errors were occurred\end{tabular} \\
invalid & \begin{tabular}[c]{@{}l@{}}GitHub could not verify the signature cryptographically\\ using the key-id attached in the signature\end{tabular} \\
malformed\_signature & Parsing error occurred while processing the signature \\
unknown\_key & \begin{tabular}[c]{@{}l@{}}The signature was created using a key that hasn't been\\linked to any user's account\end{tabular} \\
bad\_email & \begin{tabular}[c]{@{}l@{}}The email address of the committer in the commit does\\not match any of the identities associated with the PGP\\key used to create the signature\end{tabular} \\
unverified\_email & \begin{tabular}[c]{@{}l@{}}The email address of the committer is linked\\to a user, but it hasn't been verified on their account\end{tabular} \\
no\_user & \begin{tabular}[c]{@{}l@{}}The committer email address in the commit is not linked\\to any user account in GitHub\end{tabular} \\
unknown\_signature\_type & The commit contains a signature that is not PGP-based\\
unsigned & The commit is not signed \\
gpgverify\_unavailable & \begin{tabular}[c]{@{}l@{}} Unavailability of the signature verification service \end{tabular} \\
gpgverify\_error & \begin{tabular}[c]{@{}l@{}}Communication error while contacting signature\\verification service.\end{tabular} \\
not\_signing\_key & \begin{tabular}[c]{@{}l@{}}The GPG key used to create the signature does not\\include the "signing" flag in its usage settings\end{tabular} \\
expired\_key & The key used to create the signature has expired \\ \bottomrule
\end{tabular}%
}
\label{tab:signStatus}
\end{table}

\subsection{Data Analysis}

\subsubsection{\textbf{Analyzing Commit Data}}
A commit can be made directly by a user, bots, or via the GitHub web interface. GitHub automatically signs commits made by the GitHub web interface, irrespective of the users' awareness about commit-spoofing or signing. As this does not reflect commit signing practices by real users, we exclude all commits made by either GitHub's web interface or bots. Commits via the web interface can be excluded by filtering the commits made via the id ``noreply@github.com". Commits by bots are, however, difficult to detect. To find the possible bots, we find the top-$\mathit{k}$ committers (by varying $\mathit{k}$ from 5 to 20) for each repository and checked the user accounts of these committers manually to determine if it is a bot. GitHub web and bots were found to be used quite often to merge commits. Although the ``User" endpoint of GitHub's REST API can be used to check if a user account in GitHub is a ``User" or ``Bot", we found multiple instances where the fields had incorrect values. For instance, PyTorch's bot with id ``pytorchmergebot@users.noreply.github.com" is assigned a ``User" type by the REST API. Hence, we \textbf{manually} checked the possibility of an account being a bot.

\subsubsection{\textbf{Analyzing Usability of Git Clients}}
In addition to measuring the commit signing practices, we also analyzed some popular Git clients to investigate if they provide convenient features to perform commit signing and verify the integrity of commits. For this, we compared Git CLI~\cite{git}, GitHub Desktop~\cite{github-desktop}, and GitKraken~\cite{git-kraken} with GitHub's web interface~\cite{github} for the following scenarios:

\begin{itemize}
    \item \textbf{Commit signing:} Does the client provide an additional interface for setting up commit signing apart from the one provided by Git?  If yes, are they convenient to use?
    \item \textbf{Verification of commit signatures:} Does the client provide features or badges that help identify or verify a commit's signature? How interpretable are these badges, if any? 
    \item \textbf{Verification of multi-author commit signatures:} Since, only the committer can sign the commit, how do the clients show the status of such commits? How are badges or labels assigned to such commits?   
\end{itemize}

\section{Results and Findings}
\label{sec:results1}
\subsection{Prevalence and Variation in Commit Signing Usage Across Domains and Over Time}

Of the total \num{869437} downloaded commits, \num{289308} commits ($33.27\%$) were correctly signed. However, this number includes the commits made via GitHub's web interface and bots. Since commits made via GitHub's web interface are automatically signed by GitHub, and bots are mostly used for tasks like merging commits, such commits are excluded from our analysis. After excluding such commits, we were left with \num{399520} commits pushed by actual users. 

Out of \num{399520} commits pushed by actual users, only \num{38586} commits (9.65\%) were correctly signed. We further filtered the data to get unique committers in the repositories. \Cref{tab:uniqueUsers} shows the number of unique committers in each repository and the number of committers pushing signed commits. Since the table does not include the commits made via bots or via GitHub web, it only covers commits pushed directly or merged without squashing. As shown, less than 20\% of the committers have some signed or verified commits. This shows that commit signing is not well adopted, implying that the use of vigilant mode is also not adopted. We further discuss some possible reasons for the low adoption in \cref{subsubsec:low_adopt}.

\begin{figure}[h]
  \centering
  \fbox{\includegraphics[width=\linewidth]{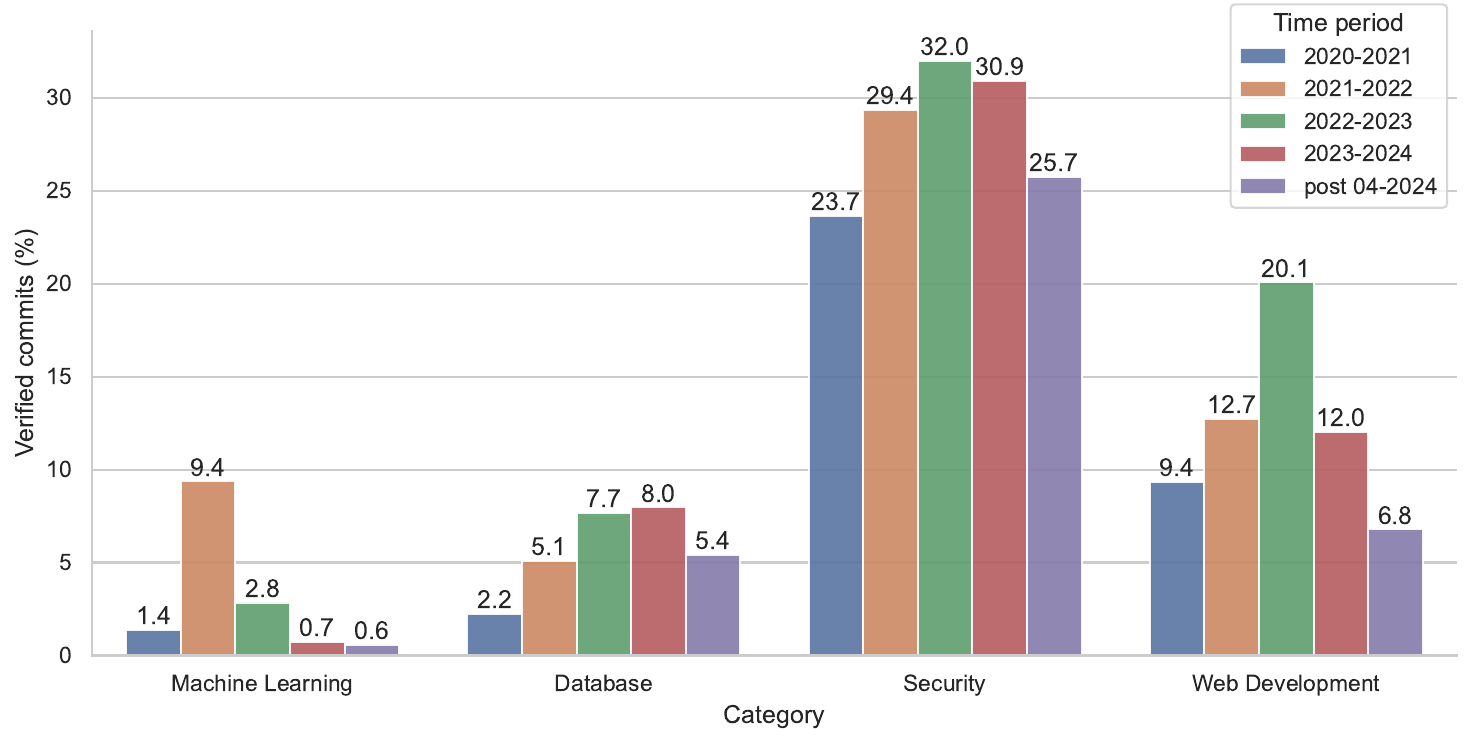}}
  \caption{Year-wise proportion of verified (valid or correctly signed) commits among different categories before and after GitHub announced vigilant mode and verification badges. This excludes commits made via bots and GitHub web. Here, in the legends, the year starts and ends on the $28^{th}$ day of April of the starting and ending year.}
  \label{fig:yearWiseProportion}
\end{figure}

We also investigated if the domain of the repositories has any effect on the usage trends of commit signing. \Cref{tab:uniqueUsers} shows that security repositories have the highest proportion of users that has some verified commits that are correctly signed followed by web development. Similar observations can be made in \cref{fig:yearWiseProportion} where the proportion of verified commits every year is the most for security repositories followed by web-development repositories. 
The numbers indicate that developers from the security domain tend to use commit signing more as compared to developers from other domains, and might be more concerned about spoofing. \Cref{tab:commitCounts} shows some insights on commit data for the top-ranked repository in each of the domains. 

\begin{table}[]
\caption{Unique committers and committers with some signed commits across the repositories studied.}
\resizebox{0.85\columnwidth}{!}{%
\label{tab:uniqueUsers}
\rowcolors{1}{white}{gray!25}
\begin{tabular}{@{}llll@{}}
\toprule
Sl No. &
  Repository Domain &
  \begin{tabular}[c]{@{}l@{}}Unique\\ Committers\end{tabular} &
  \begin{tabular}[c]{@{}l@{}}Committers with\\ correctly signed\\ commits\end{tabular} \\ \midrule
1  & Machine Learning    & 1723  & 82 (4.75\%)  \\
2  & Database        & 2504 & 160 (6.38\%) \\
3  & Security & 936 & 173 (18.48\%) \\
4  & Web development    & 1255  & 134 (10.67\%)  \\\bottomrule
\end{tabular}
}
\end{table}

\begin{table*}[]
\caption{Commit details in top-ranked repositories across all four domains with corresponding status of the commit signature.}
\resizebox{0.84\textwidth}{!}{%
\rowcolors{3}{gray!25}{white}
\begin{tabular}{@{}lllllllll@{}}
\toprule
Status
 &
  \multicolumn{2}{l}{tensorflow/tensorflow (ML)} &
  \multicolumn{2}{l}{redis/redis (DB)} &
  \multicolumn{2}{l}{fatedier/frp (Security)} &
  \multicolumn{2}{l}{facebook/react (Web-dev)} \\ \cmidrule(l){2-3} \cmidrule(l){4-5} \cmidrule(l){6-7} \cmidrule(l){8-9}
  \rowcolor{white}
 &
  Overall &
  \begin{tabular}[c]{@{}l@{}}Excluding bots\\ and GH web\end{tabular} &
  Overall &
  \begin{tabular}[c]{@{}l@{}}Excluding bots\\ and GH web\end{tabular} &
  Overall &
  \begin{tabular}[c]{@{}l@{}}Excluding bots\\ and GH web\end{tabular} &
  Overall &
  \begin{tabular}[c]{@{}l@{}}Excluding bots\\ and GH web\end{tabular} \\ \midrule
unsigned          & 86763 & 9493 & 312  & 311  & 72    & 72 & 2335 & 1252 \\
valid             & 2206  & 343  & 2784 & 1    & 344   & 0  & 4320    & 2    \\
unknown\_key      & 24    & 24   & 0    & 0    & 0     & 0   & 0     & 0    \\
no\_user          & 6     & 6    & 0    & 0    & 0     & 0   &0     & 0    \\
unverified\_email & 278   & 278  & 0    & 0    & 0     & 0   & 0     & 0    \\ \bottomrule
\end{tabular}%
}
\label{tab:commitCounts}
\end{table*}

\begin{figure*}[h]
  \centering
  \fbox{\includegraphics[width=0.84\linewidth]{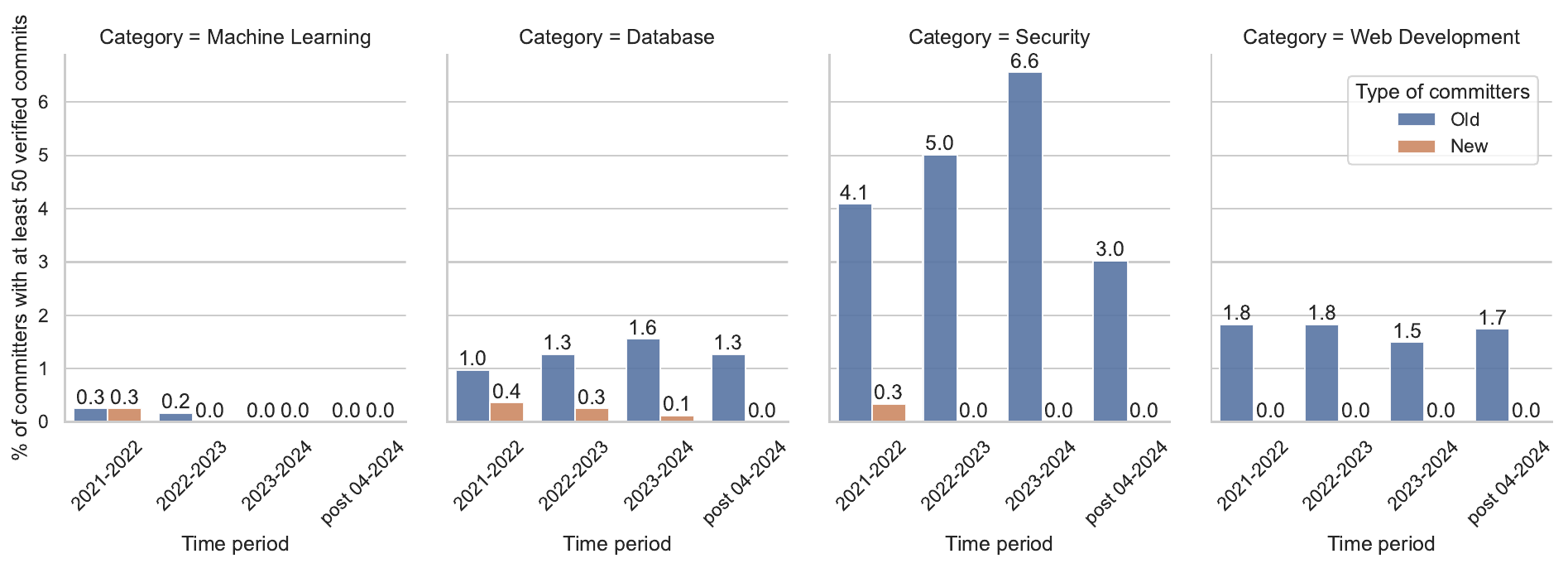}}
  \caption{Proportion of old and new committers with at least 50 verified (valid or correctly signed) commits each year. The rise and fall trend in the proportion of committers with verified commits could be the reason behind similar trend in \cref{fig:yearWiseProportion}. The drop in the proportion in the last interval could be due to the shorter range (4 months less than other intervals with 12 months).}
  \label{fig:yearWiseProportionVerifiedOldNewCommitters}
\end{figure*}

We further investigated if there is any difference between usage trends before and after GitHub released the feature to flag spoofed commits, \ie, $28^{th}$ April 2021 (we refer it as the threshold date in upcoming sections). \Cref{fig:yearWiseProportion} shows the proportion of signed commits made in each domain along with the usage trend of commit signing each year. It can be observed that for all the domains, there is a rise in the verified commits after the introduction of the verification badges (\ie, after the threshold date). For machine learning, security and web development repositories, the number of verified commits rise till early 2023 but falls after that. To identify the reason behind this, we calculated the variation of the proportion of committers committing verified commits, regularly, each year. We considered committers with at least 50 verified commits to consider users who pushed regularly. \Cref{fig:yearWiseProportionVerifiedOldNewCommitters} shows the proportion of old and new committers having verified commits out of total committers each year. The trend of committers with verified commits in \cref{fig:yearWiseProportionVerifiedOldNewCommitters} is almost similar to the trend of commits in \cref{fig:yearWiseProportion} explaining the fall in the proportion of the commits. However, the reason behind the fall in the number of committers with regular verified commits in the 2023-2024 time-frame is not clear. Another important observation here is that the number of new committers committing regularly has been decreasing, gradually.

\subsection{Usability of Commit Signing and Verification Process in Popular Git Clients}
\label{sec:result2}

Here, we assess the usability of the features provided by Git CLI, GitHub Desktop, GitHub Web and Git Kraken for commit signing and verification.

\subsubsection{\textbf{Commit Signing}} \label{subsubsec:commit_signing}
The process of commit signing via Git CLI is highly technical and procedural, involving multiple steps: (i) generate a public/private key pair, (ii) export the public key and (iii) configure Git to use the private key. Moreover, the generation of public/private key pair itself involves answering multiple questions like choosing the algorithm, key size, and expiry date. Though it is a one-time setup, it is quite rigorous and not a user-friendly option. As a CLI might not have an interactive interface, we expect GUI-based Git clients to fix this gap. In this regard, we investigated GitHub Desktop and GitKraken. We found that GitHub Desktop does not support commit signing and solely depends on Git configurations for signing commits. On the other hand, we found that GitKraken makes setting up commit signing incredibly easy by providing simple interface for generating and selecting keys. Instead of generating key pairs manually, one can simply pass the passphrase of the key to GitKraken and it automatically takes the username and email id of the user from the GitKraken profile and generates the key pair accordingly for the user. Similarly, one can select the key from the available keys via a dropdown. However, this interface is provided for GPG keys and not for SSH keys. One will have to use terminal for using SSH keys.

\subsubsection{\textbf{Verification of Commit Signatures}} For the verification of signed commits, the command \texttt{git log --show-signature} in Git CLI can be used to show signatures. However, it cannot verify the signature unless the public key is added in to the local system. Adding public key for all the users can be tedious task making the process less user-friendly. We further checked GitHub Desktop and GitKraken if they provide badges like GitHub web interface. GitHub Desktop does not provide any badge for this verification, and hence there is no graphical indication for users to check if the commit was signed. However, GitKraken provides labels specifying the status of the signature of the commits \cite{gitkraken-commit-signing}. The possible value for the labels are:
\begin{itemize}
\item GOODSIG: The key used to sign is good, and the signature has good integrity and validity.
\item EXPSIG: The signing key has expired, but the signature is still valid.
\item EXPKEYSIG: Although an expired key was used to create the signature, it is still valid.
\item REVKEYSIG: Although the signature was created with a revoked key, it is still legitimate.
\item BADSIG: The inability to verify the signature raises the possibility that there are issues with either the key or the signature itself.
\item ERRSIG: Due to errors such as unsupported algorithms or missing public keys, a signature could not be validated.
\end{itemize}

Although the labels provide detailed information about the signatures, the verification will work only if the public keys are there in the system. As discussed earlier, adding public keys can be a tedious task making the labels useless.

\subsubsection{\textbf{Verification of Multi-Authored Commit Signatures}} A commit can have multiple authors with one committer and one or more co-authors. One simple way to do this in Git is to use the command \texttt{git commit -m "Some message" --author "<uname> <email>"}. This is often used while committing via GitHub's web interface where the author is the user and committer is GitHub. However, this can also be used for spoofing. As of now, there is no way for the author to sign along with the committer and an attacker can use the identity of a victim as the author and still sign the commit. As discussed in \cref{background: badges}, GitHub uses specific badges named ``Partially Verified" for such cases. If the victim, enables vigilant mode, a correctly signed commit but with spoofed author identity will be marked as ``Partially Verified" alerting users that not all authors may have signed the commit.

Unlike GitHub's web interface, there are no badges equivalent to ``Partially Verified" in Git CLI, GitHub Desktop and GitKraken. GitHub Desktop and GitKraken simply list all the authors and committers similar to the web interface. In order to see both author and committer in Git CLI, one has to use the command \texttt{git log –format=fuller –show-signature}. However, not all users may be aware of these additional options for performing such actions.

\subsection{Improper Way of Handling Unverified Emails by GitHub}
\label{bug}

\begin{figure}[h]
  \centering
  {\includegraphics[width=0.9\linewidth]{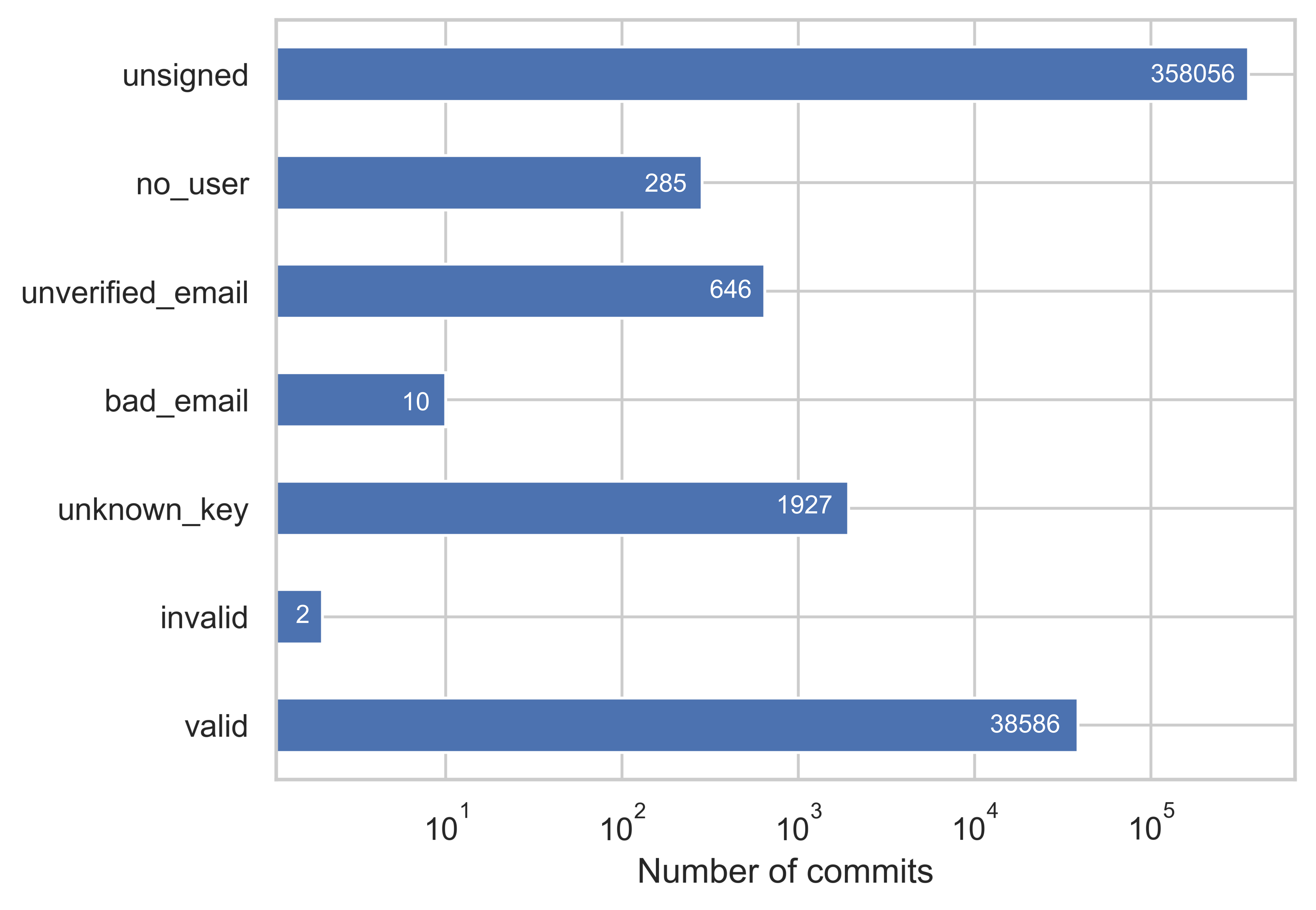}}
  \caption{Distribution of commits among different status listed in \cref{tab:signStatus}.}
  \label{fig:dist}
\end{figure}

\begin{figure*}[h]
  \centering
  \includegraphics[width=0.88\linewidth]{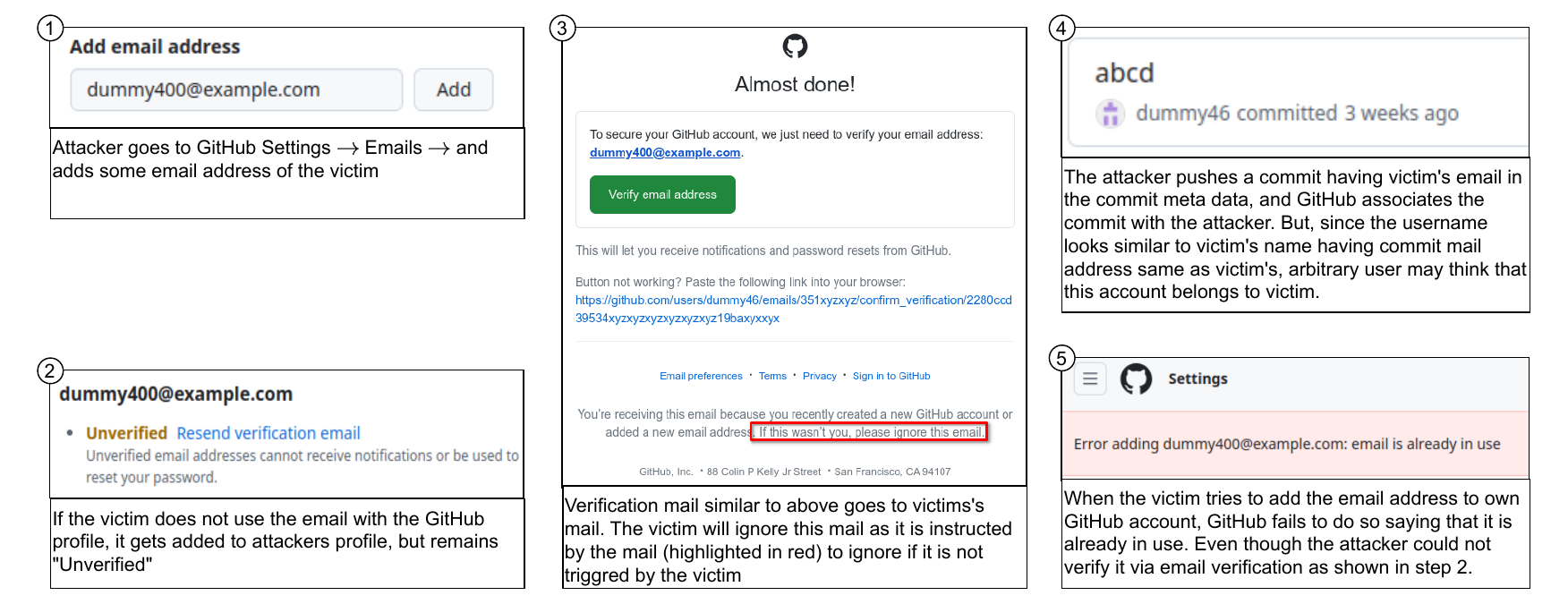}
  \caption{Email triggered when a user adds an email address to the GitHub account. If an attacker adds the email address to the account, the mail is sent to the owner. The owner has only two options: either to click on verify or ignore, as the mail clearly says (marked in red) to ignore if the owner does not trigger it. When ignored, the owner can never use the email address.}
  \label{fig:bug}
\end{figure*}

\Cref{fig:dist} shows the distribution of the commits among different status listed in \cref{tab:signStatus}, where around 646 commits had an unverified email address associated with it. An account is said to have an unverified email address if a GitHub account has added an email address to the GitHub profile but is not yet verified. These 646 commits are signed commits. The number of unsigned commits with unverified email address is unknown as the status listed in \cref{tab:signStatus} except ``unsigned" status is assigned to only those commits which were signed in some manner. Important point of concern is that even if the email address is not verified, the commits still gets associated to the user account which should not happen ideally. After further investigation, we found that an attacker can use any email address without being verified and if the owner of the email ignores the verification mail, all commits made via that email will be associated with the attacker's account. However, when the actual owner tries to add the email address to the account, GitHub prevents it from getting added with the message, ``Email is already in use" even if the email is still unverified, as shown in \cref{fig:bug}. Moreover, when the attacker adds the email address, a verification mail would be sent to the actual owner. \Cref{fig:bug} shows a sample email which gets sent when someone adds an email to a GitHub account. The recipient of the mail can either click on the ``Verify" button (verifying can lead to addition of the mail to the attackers account) or ignore the mail if its not triggered by the owner. If the owner ignores it, the owner can never use the email in its account in future.

This can lead to another kind of spoofing where an attacker can use the victim's email address to push malicious commits. Any malicious activity would then be attributed to the committer's email address which is present in the commit metadata. Ideally, if the email in the metadata of the commit is unverified, it should not be associated with any user account in GitHub. Instead, it should remain unassociated with a \texttt{no\_user} status as listed in \cref{tab:signStatus}.


\section{Discussion}
\label{sec:discussion}

\subsection{Limitations}
We discuss some limitations of our study next. Firstly, we extracted commits only from the main branch of the repositories. Sometimes commits from pull requests are squashed into one commit and merged and the original commits linked to the pull request are not present in the main branch. Our setup might have missed some commits while fetching them from the REST API. However, we were still able to analyze 399,520 commits made by real users (excluding commits made by bots and the web interface), which is a sufficiently large dataset. Secondly, we could not analyze the usage trends of the vigilant mode provided by GitHub. Vigilant mode is a significant way to tackle commit spoofing. However, there is no way of checking if a user has enabled vigilant mode via the REST API. Only way to check this is to check the verification badge assigned to a user's unsigned commit. If it is marked as unverified then the user has enabled the vigilant mode. However, to check this we need to crawl commit page of repositories, which is not allowed as per the \texttt{robots.txt} of GitHub. Finally, we plan to perform a user study in the future to study the usability and awareness of commit signing on Git clients. 

\subsection{Observations and Recommendations}
\subsubsection{\textbf{Low Adoption of Commit Signing}} \label{subsubsec:low_adopt}
As discussed above, only 9.65\% of all commits were signed correctly. Only 8.55\% of the committers in the crawled repositories had some verified commits, especially in the main or master branch of the repositories. The lower adoption implies that the users may not be aware of commit signing, including settings like the vigilant mode, which helps users prevent commit spoofing. If the users are aware but choose not to perform commit signing, the possible reasons could be:
\begin{enumerate}
    \item Though setting up commit signing is a one-time setup, the process is quite involved, as it requires knowledge of CLI tools like \texttt{ssh-keygen} and \texttt{gpg}, encryption algorithms and involves steps like generating public/private key pair, exporting and sharing public keys and configuring Git to use the keys.
    \item No recommendation by GitHub to users or enforcement by repository maintainers leading to unawareness.
    \item Commit signing needs multiple keys for different systems. Users migrating to new devices may forget to generate new keys, and managing keys across multiple devices can be challenging.
    \item Enabling vigilant mode will mark all older unsigned commits as ``Unverified''. This is something that cannot be avoided and one way to adjust with this situation would be to document the transition of the workflow.
\end{enumerate}

\subsubsection{\textbf{Usability of Git Clients for Commit Signing}}\label{disc:usability}
As discussed in \cref{subsubsec:commit_signing,subsubsec:low_adopt}, the process of setting up commit signing in Git CLI is tricky. Therefore, we need applications that provide user-friendly interface that handles the most part. For example, in GitKraken, the user needs to only input the passphrase for generating keys while all other information is taken from their GitKraken profile~\cite{gitkraken-commit-signing}. Similar interface is provided in the preferences of the applications for exporting the public key that can be shared with trusted third-parties like GitHub. However, GitHub Desktop completely relies on local Git configuration for commit signing. Having user-friendly interfaces in Git clients will encourage users to use it.
Secondly, there needs to be such badges as in GitHub web or status as in \cref{tab:signStatus}. Although GitKraken has additional labels, that works given the public key is imported in the local system. Doing this manually by a user for commits can be a tedious task. Instead, the applications can either use REST API calls to a third party like GitHub on behalf of the user to display verification status or can verify the signature by fetching the public keys on demand by users if the origin of the repository is a trusted third party like GitHub. For example, one can get the public key of a user in GitHub via a REST API call or via the URL \texttt{https://github.com/username.keys}.

In addition to this, when a repository is cloned, the Git client in use should report the distribution of verified and unverified commits, as users rarely check the commit history for verification. 

\subsubsection{\textbf{Association of User Accounts and Commits}}\label{disc:alternate}

Currently, GitHub associates user accounts with commits based on the email address in the meta data of the commit. However, associating the commits without any verification is not ideal as it leads to commit spoofing. Here, we propose an alternative way to associate commits with user accounts in GitHub. As per GitHub REST API documentation, different event gets created in GitHub based on the activity performed \cite{githubEventAPI}. GitHub has around 17 events \cite{githubEventTypes} out of which a new commit can be introduced to a repository either via a ``PushEvent" or ``PullRequestEvent". GitHub should associate a commit with a user account if the identity (username and email address) in the commit matches with identity of the user responsible for the event. If the identity does not match, it should remain unassociated with a clear message that GitHub is not sure if it associates to some account in GitHub. It is better not to associate instead of incorrectly associating it, in case of uncertainty. As GitHub already creates events during any activity performed, this process can be performed while creating such events. Although commit signing is the best way to ensure ownership and integrity, this approach is a better alternative for unsigned commits. The details of this approach in different scenarios are described below:
\begin{itemize}[leftmargin=5mm]
    \item \textbf{Push local commits to GitHub:} When a user pushes locally created commits to GitHub, all the commits created by the user would correctly get associated with the user account as the identity in the commit will match the identity of the actor in the ``PushEvent" triggered. For commits whose identity does not match with the actor of the event, GitHub can perform a check if the association between the user mentioned in the commit, and the commit with the SHA already exists in GitHub. If it exists then the pushed commit can be associated as the one that already exists otherwise it can be left unassociated. This helps when a repository is cloned and pushed as a different repository. The overhead of doing this is slightly more than searching for the user to get the public key for verifying commit signature that is already in place except that after searching the user, GitHub will also need to search the commit to check if it is already associated with the user.
    \item \textbf{Merge commits via Pull Request in GitHub:} For every action on a pull request in GitHub, a ``PullRequestEvent" is created \cite{githubEventTypes}. The action can be anyone of the opened, edited, closed, reopened, assigned, unassigned, review\_requested, review\_request\_removed, labeled, unlabeled, and synchronize. And for every such event, the commits linked to the pull request can identified via the ``commits\_url" attribute \cite{githubPRAPI}. Once the identity in the linked commits and the identity of the actor of the pull request matches, the actor can be associated with commit so that association is retained when new commits are added based on ``edit" action by other user. This way commits via pull request can be associated correctly. However, any commits made by any other committer merged locally might not get associated if the merged branch is not present in GitHub.
\end{itemize}
One major advantage of this approach is that the ownership of old commits still remains associated and verified. This is a better alternative for vigilant mode, too, because when the vigilant mode is enabled, all legitimate but unsigned commits are marked as unverified. However, our approach will ensure that the ownership of all commits are also retained. As the data transfer to GitHub happens via SSH or HTTPS, we assume that the path from the local machine to GitHub is secured as similar transfer happens for commits pushed via the web interface. Once it reaches GitHub, this approach can be used to verify ownership and associating commits.

\subsubsection{\textbf{Handling of Unverified Email in GitHub}}\label{disc:bug}
As discussed in \cref{bug}, if an email address is not already added in the profile of a user in GitHub, an attacker can add it to their own account (with a name similar to that of the actual user of the email address). The attacker cannot get it verified because the email verification will go to the actual owner. Since the verification mail asks the owner to ignore the mail as shown in \cref{fig:bug}, the owner will ignore it. However, the owner can never use the email address as GitHub does not allow this to happen, saying that it is already in use. We propose the following workflow to handle such unverified emails:

\begin{itemize}
    \item Commit with email address which is unverified should never be associated with the user account.
    \item The email verification mail should have an option to report instead of asking user to ignore. Reporting should immediately remove it from the account in which it was added.
    \item Unless the email address get verified via email verification mail, if any other user tries to add the same email address the later should be allowed to do so giving chance to the actual user. If the later is the actual owner, it will get verified and if not  the event will get reported provided the mail has an option to do so.
\end{itemize}
\section{Conclusion}

In this study, we conducted a longitudinal and cross-domain analysis of the prevalence of commit signing, focusing on real user data by excluding bot and web-based commits. Our findings indicate an increase in signed commits after GitHub introduced vigilant mode, followed by a decline around 2023-2024. Among the domains, security developers exhibited the highest usage of commit signing. Additionally, we found that GitKraken offers the most convenient commit signing experience, while GitHub web is the most accessible for verifying commits. We also identified some unexpected behavior related to unverified email handling in GitHub, which can prevent legitimate users from using their email addresses. In future, we plan to perform a rigorous user study to assess the usability and ease of commit signing among popular Git clients.


\begin{acks}
 We thank the anonymous reviewers for their feedback. This work is supported in part by the Prime Minister's Research Fellow (PMRF) grant, Science and Engineering Research Board-SRG grant, and the Google India Research Award.   
\end{acks}

\bibliographystyle{ACM-Reference-Format}
\bibliography{main}

\end{document}